\documentclass[twocolumn,superscriptaddress,prb]{revtex4-2}
\usepackage{amsfonts}
\usepackage{comment}
\usepackage{amsmath,amssymb,epsfig,color}
\usepackage{float}
\usepackage{amsmath}
\usepackage{amssymb}
\usepackage{graphicx}
\usepackage{dcolumn}
\usepackage{bbold}
\usepackage{physics}
\begin{document}

\title{Intrinsic anomalous Hall effect in altermagnets}

\author{Lotan Attias}
\affiliation{The Racah Institute of Physics, The Hebrew University of Jerusalem, Jerusalem 91904, Israel}

\author{Alex Levchenko}
\affiliation{Department of Physics, University of Wisconsin--Madison, Madison, Wisconsin 53706, USA}

\author{Maxim Khodas}
\affiliation{The Racah Institute of Physics, The Hebrew University of Jerusalem, Jerusalem 91904, Israel}

\begin{abstract}
We study the anomalous Hall effect arising from the altermagnetic order and spin-orbit interaction in doped FeSb$_2$. To investigate the anomalous transport, we have constructed a tight-binding model of FeSb$_2$. We separately considered the constraints imposed on the model parameters by the spin symmetry group and magnetic symmetry group at zero and finite spin-orbit interaction, respectively. The resulting model includes the effect of exchange splitting and is applicable at both zero and finite spin-orbit interaction.
In the case of spin symmetry, the analysis covers the spin-only subgroup arising from collinear magnetism, as well as non-trivial symmetry elements. This allows us to explore changes in the hopping amplitudes as symmetry is reduced by spin-orbit interaction from the spin group to the magnetic group. While the anomalous Hall effect is forbidden by spin symmetry, it is allowed by the symmetries of the magnetic group. The intrinsic Hall conductivity is shown to vanish linearly with spin-orbit interaction. This non-analytic behavior is universal to altermagnets. 
It originates from the singularity of the Berry curvature localized along lines on a Fermi surface confined to symmetry planes. These planes host spin degeneracy protected by spin symmetry, which is lifted by spin-orbit interaction.
\end{abstract}
\maketitle


\section{Introduction}
\label{sec:Intro}

Altermagnetism is a novel form of a collinear magnetism, distinct from both ferromagnetism and antiferromagnetism \cite{Smejkal2020,Reichlova2024,Turek2022,Smeikal2022,Mazin2022,Maier2023,Steward2023,Mazin2023,Fernandes2024}. 
For instance, in an altermagnet, zero net magnetization does not rule out a well-defined band spin polarization \cite{Mazin2021}. 
The latter arises from the combination of the antiferromagnetic order coupled to the itinerant spins and the nontrivial orbital wave function of the electronic states residing at the two magnetic sublattices, $A$ and $B$. 

The spin polarization in altermagnets is distinct from the more commonly known spin splitting induced by the spin-orbit interaction (SOI) in noncentrosymmetric materials. In the latter more familiar situation, the time-reversal symmetry ($\mathcal{T}$) is preserved while parity ($\mathcal{P}$) is broken.
In contrast, in magnets $\mathcal{T}$ is broken and the parity $\mathcal{P}$ is often preserved.

When the antiferromagnetic order causes the doubling of a unit cell, the electronic bands remain spin degenerate. 
Indeed, in such cases even though $\mathcal{T}$ is not a symmetry, the combination of $\mathcal{T}$ and a translation ($\boldsymbol{\tau}$) that exchanges the magnetic sublattices is.
As a result, spin states form the degenerate Kramers doublets related by the $\boldsymbol{\tau}\mathcal{T} \mathcal{P}$ antiunitary symmetry operation.
Similarly, if the inversion $\mathcal{P}'$ exchanging the $A$ and $B$ sublattices is a symmetry of the non-magnetic crystal the two Kramers partners are related by $\mathcal{P}'\mathcal{T}$.
In both scenarios no spin polarization of electronic bands arises. 

In contrast, magnetic order of an altermagnet does not double the unit cell, and $\mathcal{P}'$
is not a symmetry of the non-magnetic state.
To form a symmetry $\mathcal{T}$ has to be combined with some rotation. 
A given rotation, $R$ maps the generic electron momentum, $\mathbf{k}$ to another momentum $R \mathbf{k} \neq - \mathbf{k}$. 
As a consequence, a combined symmetry $R \mathcal{T}$ relates spin states at distinct momenta, $R \mathcal{T} \mathbf{k} \neq \mathbf{k}$, and therefore allows for a finite spin band splitting. 

The above spin splitting may arise from a non-trivial magnetic form factor of localized magnetic moments \cite{Hayami2018,Hayami2019,Hayami2020}.
Such moments are associated with the non-zero orbital angular momentum, $l=2,4,6$.
The exchange interaction with these moments causes the spin splitting of itinerant electrons except for the $\mathbf{k}$ along the nodal directions of the magnetisation density. 

The spin splitting described above is a hallmark of altermagnetism. 
Clearly, it is unrelated to the SOI.  
Therefore, it arises most naturally in a nonrelativistic Density Functional Theory (DFT) calculations, where the SOI is set to zero.
The symmetry of a magnetic crystal at finite SOI is defined by the specific magnetic space group of a given crystal \cite{Lifshitz2005}.
The symmetry operations in a magnetic space group act in the same way on positions of atomic sites and on the local magnetic moments.
If, e.g. such transformation is a rotation it rotates both the crystal and magnetic moments. 

\begin{figure*}[t!]
\begin{center}
\centering
\includegraphics[width=0.98\textwidth]{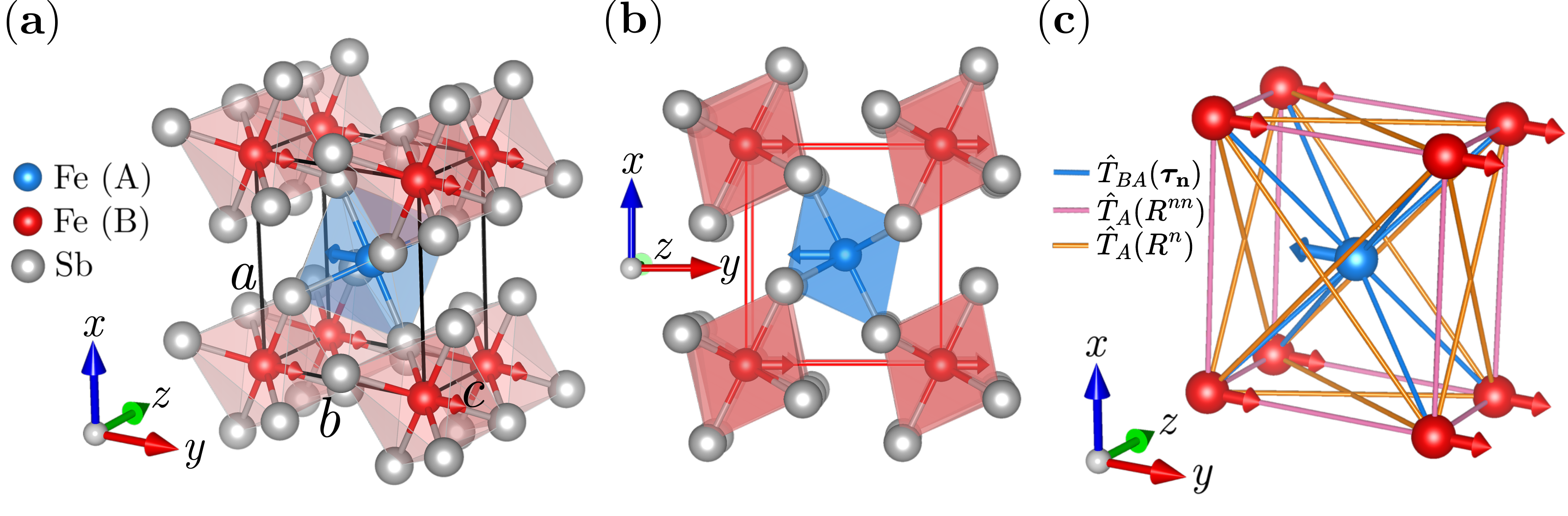}
\caption{
Crystal structure of \(\text{FeSb}_2\), space group $58$.
(a) Unit cell of \(\text{FeSb}_2\) with the cages created by Sb atoms surrounding the Fe. 
The orthorhombic lattice distances along the \( \hat{x}, \hat{y}, \hat{z} \) directions are \( a,b,c \). 
The two sublattices are oppositely polarized in the magnetic phase.
(b) YZ-plane projection of \(\text{FeSb}_2\).
(c) Tight binding hopping matrix elements for the inter-sublattice interaction, Eq.~\eqref{eq:T_BA}, and intra-sublattice interactions, Eq.~\eqref{eq:T_A(B)}. 
The three amplitudes displayed define all other hopping matrix elements. 
Presentation is aided by \textit{Vesta} \cite{VESTA_cite}. 
} \label{fig:structure_wide}
\end{center}
\end{figure*}

At zero SOI, the system may possess symmetries that act differently on spin and orbital degrees of freedom \cite{Brinkman1966}.
In this case the symmetry operations form the so called spin (space) group containing the magnetic space group as a subgroup. 
For instance, in the case of the collinear magnetism considered here such group contains a trivial, and yet important, spin-only group.
The latter contains an arbitrary angle rotation around the magnetization axis as well as rotation by $\pi$ around an axis perpendicular to the magnetization axis followed by $\mathcal{T}$.
The algorithm to construct all the possible spin groups for a given crystal and magnetic order has been formulated in Ref.~\cite{Litvin1974}.

Spin group symmetries protect extra spin degeneracies that appear as accidental from the point of view of a standard magnetic space group. 
An extended spin symmetry classification has been recently proposed to describe the spectrum degeneracies of collinear magnets in a nonrelativistic limit \cite{Smejkal2022a}. 

The observation of robust anomalous Hall effect (AHE) in altermagnets such as RuO$_2$ \cite{Smejkal2020,Feng2022} highlights the decisive role of the antiunitary symmetries in this class of magnets \cite{Hayami2021}.
Here for definiteness, we focus on the FeSb$_2$ with an orthorhombic unit cell containing two distinct Fe atoms located at the $A$ and $B$ sublattices respectively. 
Fe atoms at the $A$ and $B$ sublattices are oppositely spin-polarized along $\hat{y}$, see Fig.~\ref{fig:structure_wide}a,b.

The symmetry operations are outlined in Sec.~\ref{sec:Symmetries}.
In Sec.~\ref{sec:TB} we construct the minimal tight-binding model at zero and finite SOI.
The AHE is studied in Sec.~\ref{sec:AHE}.
We summarize in Sec.~\ref{sec:Summary}. 

\section{Symmetries and band degeneracies}
\label{sec:Symmetries}
In this section we introduce the magnetic space group $Pnn'm'$ appropriate to FeSb$_2$ with finite SOI.
We list the elements of the magnetic point group, $\mathcal{G}_M = G_M/\mathbf{R}$, which is a factor group of magnetic space group, $G_M$ by the group of translations, $\mathbf{R}$.
$\mathcal{G}_M$ contains unitary and antiunitary elements, and we employ the Wigner criterion to find out the momenta with the double degeneracy due to the antiunitary symmetries.

We then repeat the above steps for the spin group of anti-ferromagnetic FeSb$_2$ with zero SOI.
The degeneracies have been studied in Ref.~\cite{Mazin2021}.
We rely on the results below as a consistency check.
Moreover, it is of interest to apply the Wigner trichotomy test in a non-standard setting of spin groups.

\subsection{Finite SOI}

In this case the magnetic point group can be presented as 
$\mathcal{G}_M = \mathcal{G}^u_M+\mathcal{T}C_{2z}\mathcal{G}^u_M$.
Here the unitary elements form a simple abelian subgroup, 
$\mathcal{G}^u_M = \{ E, \mathcal{P} , \boldsymbol{\tau} C_{2x}, \boldsymbol{\tau} m_x \}$,
where $E$ is identity operation, $m_x = \mathcal{P}C_{2x}$ is a mirror in $yz$-plane. 

In the case of $\mathcal{G}_M$, the Wigner criterion states that for a given $\mathbf{k}$ one computes the sum to distinguish between the three cases,
\begin{align}\label{eq:W_test}
    \sum'_{g \in \mathcal{G}^u_M}\chi^{\mathbf{k}}\left[ (C_{2z} g)^2 \right] = \begin{cases}
    + [\mathbf{k}]\mathcal{T}^2  & \mathrm{case \,(a)} \\
    -[\mathbf{k}]\mathcal{T}^2  & \mathrm{case \,(b)} \\
    0 & \mathrm{case \,(c)}
    \end{cases}\, .
\end{align}
In the case (a) the antiunitary operation, $C_{2z} \mathcal{T}$ does not cause the extra degeneracy, and in cases (b) and (c) it does. 
The band degeneracy, in addition, requires that there are elements $g \in \mathcal{G}_u$ such that $C_{2z} g$ reverses $\mathbf{k}$ \cite{Inui1990,Liao2012}.
The summation in Eq.~\eqref{eq:W_test} runs over $[\mathbf{k}]$ such elements. 
The $\chi^{\mathbf{k}}$ is the irreducible character of the group of $\mathbf{k}$, $G_{\mathbf{k}}$.

Consider for illustration the generic point in the $k_y = \pi$ plane, $\mathbf{k}=(k_x,\pi,k_z)$.
For generic $k_x$ and $k_z$, the group $\mathcal{G}_{\mathbf{k}}$ contains only translations, $\mathbf{R}$.
The sum in Eq.~\eqref{eq:W_test} reduces to a single term, $g = \boldsymbol{\tau} C_{2x}$, $[\mathbf{k}]=1$.
And since this operation squares to $(C_{2z}\boldsymbol{\tau} C_{2x})^2 = - \mathbf{R}_yC_{2y}^2 $ with $\mathbf{R}_y = \hat{y}$, the sum in the Wigner criterion is trivially evaluated,
\begin{align}\label{eq:W_test_res}
    \chi^{\mathbf{k}}\left[ (C_{2z} g)^2 \right] = - \mathcal{T}^2\, .
\end{align}
Indeed, $C_{2y}^2 = \mathcal{T}^2= \mp 1$ for (half)-integer spin, and according to Bloch theorem, $\chi^{\mathbf{k}}(\mathbf{R}) = \exp(- i \mathbf{k} \mathbf{R})$.
Based on Eq.~\eqref{eq:W_test_res} we conclude that the degeneracy of Bloch bands at $k_y = \pi$ plane doubles due to the $C_{2z} \mathcal{T}$ antiunitary symmetry. Similar analysis shows that this statement holds true for two lines parallel to the $\hat{y}$ axis, ($k_x=0, k_z = \pi$) and ($k_x=\pi, k_z = 0$).
In total, the subset with the double degeneracy, $\mathcal{K}_{\mathrm{SO}}$ included one plane and two lines.

\subsection{Zero SOI}
\label{sec:Zero_SOI}
As discussed in Sec.~\ref{sec:Intro}, this case requires an analysis of the spin group, $G_S$. 
Similar to magnetic groups, the spin point group, $\mathcal{G}_S = G_S/\mathbf{R}$ contains the unitary elements $\mathcal{G}^u_S$ as a subgroup of index two,
\begin{align}\label{eq:SG_13}
   \mathcal{G}^u_S = & \left[ ( C_{\infty y} || E) + \boldsymbol{\tau}C_{2x} ( C_{\infty y} || E)  \right]
 \notag \\
& \times  \left\{ (E || E ) , (E || \mathcal{P} ), (E || m_z ) ,  (E || C_{2z} ) \right\}\, ,
\end{align}
where $g(g_s||g_o)$ denotes the operation with $g_s$ ($g_o$) acting on spin (orbital) degrees of freedom, respectively followed by the unitary or anti-unitary operation $g$ acting in the same way on both degrees of freedom. 
Operations $( C_{\infty y} || E)$ form a unitary subgroup of the spin-only group,
$( C_{\infty y} || E) + \mathcal{T}( C_{2x} || E) ( C_{\infty y} || E)$, acting only on spins \cite{Litvin1974}. 
It emerges as spins decoupled from the orbital motion can be freely rotated around the magnetization axis by an arbitrary angle without affecting the Hamiltonian. 
The second line of Eq.~\eqref{eq:SG_13} reflects the $C_{2h}$ site symmetry of the Sb cage enclosing Fe atoms at both sublattices. 

The full spin point group including the anti-unitary operations can then be represented in a compact form as
\begin{align}\label{SpG_60}
    \mathcal{G}_S = \mathcal{G}^u_S + \boldsymbol{\tau} \mathcal{T} (E || C_{2x} )\mathcal{G}^u_S\, .
\end{align}
Before considering the possible doubling of the degeneracy due to non-unitary symmetries, one has to determine the irreducible representation of the group of $\mathbf{k}$, $G_{\mathbf{k}}$.
The elements acting on the orbital degrees of freedom form a $D_{2h}$ abelian group.
This allows us to focus on spin degeneracy.

\begin{table}
\caption{\label{tab:SG1} 
The action of the orbital part of the operations from the list $C_{2x} \left\{ E  , \mathcal{P} , m_z , C_{2z}  \right\} $ on the momentum $\mathbf{k}=(k_x,k_y,k_z)$.}
\begin{tabular}{|c|c|c|c|}
\hline
$C_{2x}$ & $m_x$ & $m_y$ & $C_{2y}$ 
 \\
 \hline
 $(k_x, -k_y, -k_z)$ & $(-k_x, k_y, k_z)$   & $(k_x, -k_y, k_z)$   & $(-k_x, k_y, -k_z)$  
\\
\hline
\end{tabular}
\end{table}

At a generic $\mathbf{k}$, up to translations, $G_{\mathbf{k}} = ( C_{\infty y} || E) $ and the irreducible representations are one dimensional. 
Hence there is no spin degeneracy unless $\mathbf{k}$ is invariant under one of the four operations,
$C_{2x} \left\{ E  , \mathcal{P} , m_z , C_{2z}  \right\}$.
These operation are listed in Tab.~\ref{tab:SG1} along with their action on $\mathbf{k}$.
It follows directly from Tab.~\ref{tab:SG1} that the spin degeneracy is doubled at four planes $k_x=0,\pi$ and $k_y=0,\pi$ comprising the set $\mathcal{K}_{\mathrm{alt}}$.
For instance, $k_y=\pi$ plane is invariant under $C_{2x} m_z = m_y$ operation.

In contrast to the magnetic group, the anti-unitary operations of the spin group do not lead to the degeneracy doubling.
We demonstrate this by extending the Wigner criterion to the spin group (see App.~\ref{SI:A} for details).


\section{Single orbital tight-binding model}
\label{sec:TB}

We construct the tight-binding model of itinerant electrons to incorporate the SOI and the exchange coupling in an altermagnet. 
We start with the atomic limit by looking at the effect of the lattice on electronic states localized at a given lattice site. The exchange splitting, $2 B_{ex}$ is assumed to be much larger than the spin splitting induced by the SOI locally at a given site. This makes it reasonable to ignore the local effect of SOI.
In contrast, we do study in detail the effect of SOI on the hopping amplitudes to the neighboring sites in the following sections.

In the absence of the local SOI, the spin and orbital degrees of freedom at each site decouple.
We therefore, discuss the localized orbital wave-functions ignoring the spin. 
The site symmetry group is abelian, and Sb cage lifts the orbital degeneracy at Fe sites.
Hence, we focus on a single orbital model with orbital wave-functions $\phi_{A(B)}(\mathbf{r})$ at the $A$ and $B$ sublattices.

\begin{figure}
\begin{center}
\centering
\includegraphics[width=0.48\textwidth]{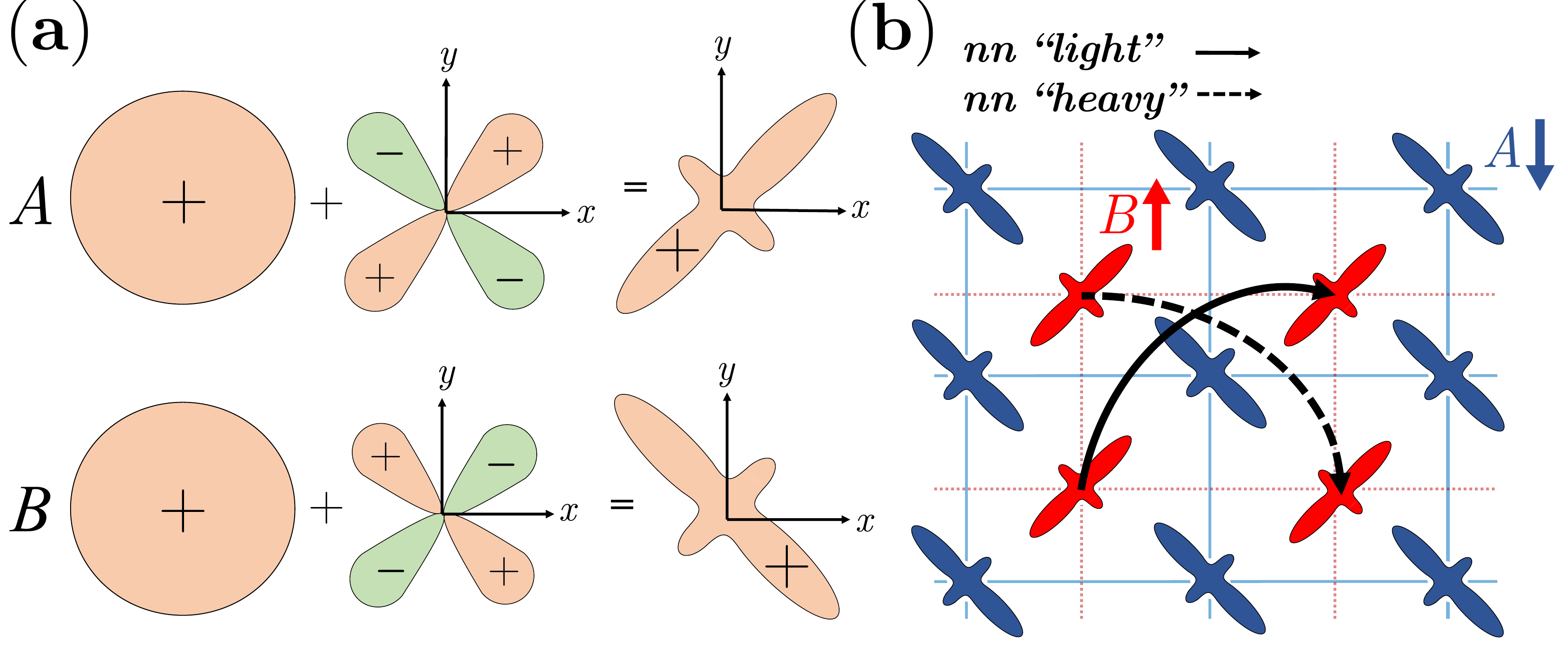}
\caption{
(a) Symmetry-allowed hybridization of $s$ and $d_{xy}$ orbitals at Fe sites.
The hybridized wave functions at the $A$ and $B$ sublattices are related by the nonsymmorphic $\boldsymbol{\tau}C_{2x}$ symmetry operation.
As a result, the hybridized orbital wave function at the two sublattices are distinct. 
(b) 
The projection of the crystal structure on the $xy$-plane.
The difference in the orbital wave functions for the two sublattices implies $A' \neq 0$ in Eq.~\eqref{eq:Ha_nn} causing a spin splitting even at zero SOI. 
} \label{fig:orbitals}
\end{center}
\end{figure}

The nonsymmorphic $\boldsymbol{\tau}C_{2x}$ symmetry implies the relationship $\phi_{B}(\mathbf{r}) = \phi_{A}(C_{2x}^{-1}\mathbf{r})$.
It is crucial for altermagnetism that the functional form of the $\phi_A(\mathbf{r})$ and $\phi_B(\mathbf{r})$ orbital wave functions of the two members in each degenerate doublet may differ.
Indeed the on-site symmetry allows orbitals of same parity that are both even or odd with respect to $C_{2z}$ to hybridize. 
This makes room for the hybridization of orbitals transforming differently under $C_{2x}$.
For instance, the hybridization of $s$- and $d_{xy}$-orbitals gives rise to distinct orbital wave functions at the two sublatices, see Fig.~\ref{fig:orbitals}a.
This point is further elaborated upon in Sec.~\ref{sec:Intra}.

These considerations naturally lead us to the tight-binding four band model
\begin{align}\label{eq:H_TB}
    \mathcal{H} = \sum_{\mathbf{k}} \Psi_{\mathbf{k} \alpha}^{\dagger} \hat{H}_{\alpha \beta}( \mathbf{k})\Psi_{\mathbf{k} \beta}\, 
\end{align}
expressed in terms of the generalized Nambu spinor,
\begin{align}\label{eq:Psi}
  \Psi^\dagger_{\mathbf{k}} = \left[ \psi^\dagger_{\mathbf{k}A \uparrow}, \psi^\dagger_{\mathbf{k}A\downarrow}, \psi^\dagger_{\mathbf{k}B\uparrow}, \psi^\dagger_{\mathbf{k}B\downarrow}\right]\, .  
\end{align}
When applied to vacuum, $|0 \rangle$, $\psi^\dagger_{\mathbf{k}As}$ and $\psi^\dagger_{\mathbf{k}Bs}$ create the electronic Bloch states localized at the $A$- and $B$-sublattices respectively, with two possible spin projections on the $\hat{z}$ axis.
Specifically, 
\begin{subequations}\label{eq:wf}
\begin{align}\label{eq:wfA}
    \psi^\dagger_{\mathbf{k}A s}|0\rangle & =
    \frac{1}{\sqrt{N}}\sum_{\mathbf{R} \in A} e^{ i \mathbf{k} \mathbf{R}} \phi_{A}(\mathbf{r} - \mathbf{R} )\chi_s \, , 
\end{align}    
\begin{align}\label{eq:wfB}
    \psi^\dagger_{\mathbf{k}B s}|0\rangle & =
    \frac{1}{\sqrt{N}}
    \sum_{\mathbf{R} \in A} e^{ i \mathbf{k} (\mathbf{R} + \boldsymbol{\tau})} \phi_{B}(\mathbf{r} - \mathbf{R} - \boldsymbol{\tau} )\chi_s\, ,
\end{align}  
\end{subequations}
where $N$ Fe ions of the $A$-sublattice are located at the sites of the orthorhombic lattice, $\mathbf{R} = n_x a \hat{x} + n_y b \hat{y} + n_z c\hat{z}$ defined by the vector, 
$\mathbf{n}=(n_x,n_y,n_z)$ with integer valued components, see Fig.~\ref{fig:structure_wide}a.
Hereinafter, the distances along the three crystallographic directions are measured in units of $a$, $b$ and $c$, respectively.
Correspondingly, the momentum components $k_{x}$, $k_y$ and $k_z$ are measured in units of $a^{-1}$, $b^{-1}$ and $c^{-1}$.

To parametrize the Hamiltonian \eqref{eq:H_TB} we introduce the two sets of Pauli matrices, $\boldsymbol{\varkappa}$ and $\boldsymbol{\sigma}$ acting in sublattice and spin spaces, respectively.
The unit matrices acting in these spaces are denoted as $\varkappa_0$ and $\sigma_0$.
The generic model Hamiltonian reads,
\begin{align}\label{eq:Htotal}
    \hat{H}(\mathbf{k}) = \hat{H}^e(\mathbf{k}) +\hat{H}^a_n(\mathbf{k})  + \hat{H}^a_{nn}(\mathbf{k}) + \hat{H}_{ex}(\mathbf{k})\, ,
\end{align}
where the exchange interaction is $\hat{H}_{ex} = B_{ex} \varkappa_z \sigma_y$, $\hat{H}^e(\mathbf{k})$ is the contribution of the nearest neighbour inter-sublattice hopping processes, $\hat{H}^a_n(\mathbf{k})$ and  
$\hat{H}^a_{nn}(\mathbf{k})$ originate from the nearest and next to nearest neighbor intra-sublattice hopping processes, respectively. 
Below we obtain the generic form of these terms constrained by the $\mathcal{G}_M$ and $\mathcal{G}_S$ symmetries.


\subsection{Nearest neighbor inter-sublattice hopping processes}
\label{sec:NNinter}

We start with the analysis of $\hat{H}^e(\mathbf{k})$ part of the tight-binding Hamiltonian. 
The real space hopping matrix elements of the full microscopic space periodic Hamiltonian, $H$ are introduced in a standard way,
\begin{align}\label{eq:T_BA}
    T^{s' s}_{BA}(\boldsymbol{\tau}_{\mathbf{n}}) = \langle \phi_{B}( \mathbf{r} - \boldsymbol{\tau}_{\mathbf{n}}) \chi_{s'} | H |\phi_{A}(\mathbf{r}) \chi_{s}  \rangle\, .
\end{align}
A given $A$ site has eight neighboring sites to hop to in the $B$-sublattice with the hopping vectors, $\boldsymbol{\tau}_{\mathbf{n}} = (1/2)( n_x \hat{x} + n_y \hat{y} + n_z \hat{z})$ fixed by the choice of $n_{x,y,z} = \pm 1$, see Fig.~\ref{fig:structure_wide}c. 
In particular, $\boldsymbol{\tau}_{\mathbf{n}} = \boldsymbol{\tau}$ for $n_x=n_y=n_z = 1$.
It is convenient to represent the spin dependence of amplitude Eq. \eqref{eq:T_BA} in the form,
\begin{align}\label{eq:T_BA_1}
    \hat{T}_{BA}(\boldsymbol{\tau}_{\mathbf{n}}) = 
    \sum_{\mu} t_\mu(\boldsymbol{\tau}_{\mathbf{n}}) \sigma_\mu ,
\end{align}
where $\mu$ runs over the four values, $0,x,y,z$.

The tight-binding Hamiltonian resulting from the inter-sublattice hopping processes,
\begin{align}\label{eq:H_e_k}
    \hat{H}_{e}(\mathbf{k}) = \sum_\mu \left[ \varkappa_x t^R_{\mu}(\mathbf{k}) + \varkappa_y t^I_{\mu}(\mathbf{k}) \right] \sigma_\mu\, ,
\end{align}
where $t^R_{\mu}(\mathbf{k})$ and $t^I_{\mu}(\mathbf{k})$ are real and imaginary parts of 
\begin{align}\label{eq:t_k}
    t_{\mu}(\mathbf{k}) =
    \sum_{\boldsymbol{\tau}_{\mathbf{n}}} e^{- i \mathbf{k}[ \Delta \mathbf{R}(\boldsymbol{\tau}_{\mathbf{n}})+\boldsymbol{\tau}]}  
    t_{\mu}(\boldsymbol{\tau}_{\mathbf{n}})\, .
\end{align}
Here $\Delta \mathbf{R}(\boldsymbol{\tau}_{\mathbf{n}})$ are the Bravais lattice vector connecting the unit cells hosting the neighboring sites at $A$ and $B$ sublattices, see Tab.~\ref{tab:t}. 

\subsubsection{Finite SOI}

\begin{table}
\caption{\label{tab:t}
The nearest neighbor inter-sublattice hopping parameters constrained by Eq.~\eqref{eq:t_constr}.
The four out of eight hopping vectors, $\boldsymbol{\tau}_{\mathbf{n}} =\mathbf{n}/2$ are sufficient in view of Eq.~\eqref{eq:P}, and $\boldsymbol{\tau}_{-\mathbf{n}} = - \boldsymbol{\tau}_{\mathbf{n}}$.
The Bravais lattice vector, $\Delta \mathbf{R}(\boldsymbol{\tau}_{\mathbf{n}})$ is introduced in Eq.~\eqref{eq:t_k}.
For the neighboring sites that are not shown it is given by,
$\Delta \mathbf{R}(-\boldsymbol{\tau}_{\mathbf{n}}) =
-\hat{x}-\hat{y}-\hat{z} - \Delta \mathbf{R}(\boldsymbol{\tau}_{\mathbf{n}})$. These parameters are visualized in Fig.~\ref{fig:hopping_amp}a of App.~\ref{SI:amplitudes}.}
\begin{tabular}{c|c|c|c|c|} 
\hline 
   $\mathbf{n}$ & $(1,1,1)$ & $(-1,1,1)$ & $(1,-1,1)$ & $(1,1,-1)$   \\
    \hline
     $t_{0}(\boldsymbol{\tau}_{\mathbf{n}})$ & $t_{0}$ & $t^*_{0}$ & $t_{0}$ & $t^*_{0}$  \\
    $t_{x}(\boldsymbol{\tau}_{\mathbf{n}})$ & $t_{x}$ & $t^*_{x}$ & $t_{x}$ & $t^*_{x}$  \\
    $t_{y}(\boldsymbol{\tau}_{\mathbf{n}})$ & $t_{y}$ & $-t^*_{y}$ & $-t_{y}$ & $t^*_{y}$ \\
    $t_{z}(\boldsymbol{\tau}_{\mathbf{n}})$ & $t_{z}$ & $-t^*_{z}$ & $t_{z}$ & $-t^*_{z}$ \\
    \hline
    $\Delta \mathbf{R} (\boldsymbol{\tau}_{\mathbf{n}})$ & $0$ & $-\hat{x}$ & $-\hat{y}$ & $-\hat{z}$  \\
    \hline 
\end{tabular}
\end{table}

We now turn to the constrains imposed on the matrix elements Eq.~\eqref{eq:T_BA} by the magnetic  point group, $\mathcal{G}_M$. 
The parity $\mathcal{P}$ imposes the condition,
\begin{align}\label{eq:P}
     \hat{T}_{BA}(\boldsymbol{\tau}_{\mathbf{n}})=\hat{T}_{BA}(-\boldsymbol{\tau}_{\mathbf{n}})\, .
\end{align}
The unitary symmetry $\boldsymbol{\tau}C_{2x}$ imposes the constrain, 
\begin{align}\label{eq:tauC2x}
 \hat{T}_{BA}(\boldsymbol{\tau}_{\mathbf{n}}) =\sigma_x [\hat{T}_{BA}(C_{2x} \boldsymbol{\tau}_{\mathbf{n}}) ]^\dagger \sigma_x\, .
\end{align}
The antiunitary symmetry $C_{2z}\mathcal{T}$ leads to the condition,
\begin{align}\label{eq:TC2z}
    \hat{T}_{BA}(\boldsymbol{\tau}_{\mathbf{n}}) = 
     \sigma_x [\hat{T}_{BA}(C_{2z} \boldsymbol{\tau}_{\mathbf{n}}) ]^* \sigma_x\, .
\end{align}
Applying the set of the three equations \eqref{eq:P}, \eqref{eq:tauC2x} and \eqref{eq:TC2z} to Eq.~\eqref{eq:T_BA_1} yields the following constrains, 
\begin{subequations}\label{eq:t_constr}
    \begin{align}
    t_{0(x)}(\boldsymbol{\tau}_{\mathbf{n}})\! =\! t_{0(x)}(-\boldsymbol{\tau}_{\mathbf{n}}) \! =\! t^*_{0(x)}(C_{2x} \boldsymbol{\tau}_{\mathbf{n}}) \! =\! t^*_{0(x)}(C_{2z} \boldsymbol{\tau}_{\mathbf{n}}),
\end{align}
\begin{align}
t_{y}(\boldsymbol{\tau}_{\mathbf{n}})  = t_{y}(-\boldsymbol{\tau}_{\mathbf{n}}) = -t^*_{y}(C_{2x} \boldsymbol{\tau}_{\mathbf{n}}) = t^*_{y}(C_{2z} \boldsymbol{\tau}_{\mathbf{n}})\, ,
\end{align}
\begin{align}
    t_{z}(\boldsymbol{\tau}_{\mathbf{n}})  = t_{z}(-\boldsymbol{\tau}_{\mathbf{n}}) = -t^*_{z}(C_{2x} \boldsymbol{\tau}_{\mathbf{n}}) = -t^*_{z}(C_{2z} \boldsymbol{\tau}_{\mathbf{n}})\, .
\end{align}
\end{subequations}
Equation \eqref{eq:t_constr} implies that the $\mathcal{G}_M$ reduces the 32 complex parameters in Eq.~\eqref{eq:T_BA} down to 4, $t_\mu = t_\mu(\boldsymbol{\tau})$.
The 8 real parameters fixing the Hamiltonian Eq.~\eqref{eq:H_e_k} are the real and imaginary parts, $t^R_\mu$ and $t^I_\mu$ of $t_\mu$.
The rest of the hopping amplitudes follow from Eq.~\eqref{eq:t_constr} as summarized in Tab.~\ref{tab:t}.
It turns out to be convenient to split the resulting tight-binding Hamiltonian Eq.~\eqref{eq:H_e_k} into two parts,
$\hat{H}^{e}(\mathbf{k}) = \hat{H}_{\mathcal{T}}^{e}(\mathbf{k}) + \hat{H}_{\mathrm{B}}^{e}(\mathbf{k})$, where the first part $\hat{H}_{\mathcal{T}}^{e}(\mathbf{k})$ is invariant under $\mathcal{T}$, and the second part $\hat{H}_{\mathrm{B}}^{e}(\mathbf{k})$ breaks $\mathcal{T}$ and is associated, therefore with the exchange field, $\mathbf{B}$. 
We have
\begin{align}\label{eq:HeT}
  \hat{H}_{\mathcal{T}}^{e}(\mathbf{k})&  =   
  8 t_0^R \varkappa_x \sigma_0 \cos\frac{k_x}{2} \cos\frac{k_y}{2} \cos\frac{k_z}{2}  
  \notag \\
  - &   8 t_x^I \varkappa_y \sigma_x \sin\frac{k_x}{2} \cos\frac{k_y}{2} \sin\frac{k_z}{2}  
   \notag \\
  - &   8 t_y^I \varkappa_y \sigma_y \cos\frac{k_x}{2} \sin\frac{k_y}{2} \sin\frac{k_z}{2} 
   \notag \\
  + &   8 t_z^I \varkappa_y \sigma_z \cos\frac{k_x}{2} \cos\frac{k_y}{2} \cos\frac{k_z}{2} \, ,
\end{align}
and the terms induced by the exchange field, 
\begin{align}\label{eq:HeB}
  \hat{H}_{\mathrm{B}}^{e}(\mathbf{k})&  =   
  - 8 t_0^I \varkappa_y \sigma_0 \sin\frac{k_x}{2} \cos\frac{k_y}{2} \sin\frac{k_z}{2}  
  \notag \\
   &   8 t_x^R \varkappa_x \sigma_x \cos\frac{k_x}{2} \cos\frac{k_y}{2} \cos\frac{k_z}{2}  
   \notag \\
  - &   8 t_y^R \varkappa_x \sigma_y \sin\frac{k_x}{2} \sin\frac{k_y}{2} \cos\frac{k_z}{2} 
   \notag \\
  - &   8 t_z^R \varkappa_x \sigma_z \sin\frac{k_x}{2} \cos\frac{k_y}{2} \sin\frac{k_z}{2} \, .
\end{align}

The spectrum of both Eqs.~\eqref{eq:HeT} and \eqref{eq:HeB} taken separately is Kramers degenerate.
In the case of Eq.~\eqref{eq:HeT} this is due to the combined $\mathcal{T}\mathcal{P}$ symmetry. 
In the case of Eq.~\eqref{eq:HeB} it occurs because of the $\varkappa_z \mathcal{T}\mathcal{P}$ thanks to the chiral symmetry, $\varkappa_z$ characteristic for the inter-sublattice processes on a bipartite lattice.

The terms respecting $\mathcal{T}$ symmetry, Eq.~\eqref{eq:HeT}, are obtained when the exchange field breaking $\mathcal{T}$ is set to zero. 
They, therefore give the usual SOI, and as expected agree with the results of Ref.~\cite{Roig2024}, where the $\mathcal{T}$-breaking term describing the coupling to the exchange field is added separately.

In contrast, the terms breaking $\mathcal{T}$ symmetry, Eq.~\eqref{eq:HeB}, require a finite $\mathbf{B}$.
Some of these terms result from the combined action of the SOI and the exchange field.
Still, others are independent of SOI. 
The framework of the spin symmetries allows one to disentangle these two spin dependent interactions.
And that is what we do next.

\begin{table}
\caption{\label{tab:tnnn}
The next nearest neighbor intra-sublattice hopping parameters constrained by the symmetry.
The six out of twelve amplitudes are shown.
The rest follows from the symmetry of all the amplitudes under reversal of $\Delta\mathbf{R}$. These parameters are visualized in Fig.~\ref{fig:hopping_amp}b of App.~\ref{SI:amplitudes}.}
\begin{tabular}{c|c|c|c|c|c|c|} 
\hline 
   $\Delta\mathbf{R}$ & $\hat{x}+\hat{y}$ & $\hat{x}-\hat{y}$ & $\hat{y}+\hat{z}$ & $\hat{y}-\hat{z}$  & $\hat{z}+\hat{x}$ & $\hat{z}-\hat{x}$ \\
    \hline
    $T_{A}(\Delta\mathbf{R})$ & $A_+$ & $A_-$ & $A_1$ & $A_1$ & $A_2$ & $A_2$  \\
    $T_{B}(\Delta\mathbf{R})$ & $A_-$ & $A_+$ & $A_1$ & $A_1$ & $A_2$ & $A_2$\\
    \hline 
\end{tabular}
\end{table}

\subsubsection{Zero SOI}

Since the magnetic group $\mathcal{G}_M$ is the subgroup of the spin group, $\mathcal{G}_S$, Eqs.~\eqref{eq:HeT} and \eqref{eq:HeB} still hold, possibly with some of the coefficients forced to be zero due to a larger spin symmetry group.
The spin-only group implies that all the terms in $\hat{H}_{e}(\mathbf{k})$ proportional to $\sigma_x$ or $\sigma_z$ are zero. 
In particular, we have $t^{R,I}_x = t^{R,I}_z=0$ in Eqs.~\eqref{eq:HeT} and \eqref{eq:HeB}.
This conclusion is quite obvious given that the magnetism in the case of zero SOI is collinear.

The less obvious conclusions are obtained as one considers a non-trivial spin group operations.
The $(E||m_z)$ operation acting on orbital degrees of freedom implies 
\begin{align}\label{eq:TSG13}
    \hat{T}_{BA}(\boldsymbol{\tau}_{\mathbf{n}}) = 
     \hat{T}_{BA}(m_z \boldsymbol{\tau}_{\mathbf{n}})\, ,
\end{align}
which upon comparison with Tab.~\ref{tab:t} yields $t_{0} = t^*_{0}$ and $t_{y} = t^*_{y}$.
The non-unitary symmetry $\boldsymbol{\tau} \mathcal{T} (E|| C_{2x})$ yields the constrain,
\begin{align}
   \hat{T}_{BA}(\boldsymbol{\tau}_{\mathbf{n}}) = \sigma_y \hat{T}_{BA}^{tr}( C_{2x} \boldsymbol{\tau}_{\mathbf{n}})\sigma_y\, .
\end{align}
This condition does not constrain $\hat{H}_{e}(\mathbf{k})$ farther. 

The rest of the operations beyond the spin-only group are obtained by combining the operations considered so far with $\mathcal{G}_M$, and therefore do not result in additional constrains.
As a consistency check we may consider the operation, $\mathcal{T} (C_{2x}||E) = (\boldsymbol{\tau}C_{2x} )\boldsymbol{\tau} \mathcal{T} (E|| C_{2x})$.
It implies 
\begin{align}
  \hat{T}_{BA}(\boldsymbol{\tau}_{\mathbf{n}}) = \sigma_z  \left[\hat{T}_{BA}(\boldsymbol{\tau}_{\mathbf{n}})\right]^* \sigma_z\, .
\end{align}
And again, $t_0^I = t_x^I =0$.
As expected, it contains no new information.
In summary, at zero SOI and finite exchange field, the inter-sublattice hopping Hamiltonian takes the form,
\begin{align}\label{eq:He_noSO}
  \hat{H}^e(\mathbf{k})&  =   
  8 t_0^R \varkappa_x \sigma_0 \cos\frac{k_x}{2} \cos\frac{k_y}{2} \cos\frac{k_z}{2}  
    \notag \\
  - &   8 t_y^R \varkappa_x \sigma_y \sin\frac{k_x}{2} \sin\frac{k_y}{2} \cos\frac{k_z}{2} \, .
\end{align}
Here the first term is the usual spin independent contribution to the dispersion.
The second term is the spin dependent contribution arising ultimately from the exchange interaction.
Although this term is similar in form to the SOI, it is fundamentally distinct from it.
It breaks $\mathcal{T}$ and exists at zero SOI interaction. 

\subsection{Intra-sublattice hopping processes}
\label{sec:Intra}
We now turn to the intra-sublattice part of the tight-binding Hamiltonian, $\hat{H}^a(\mathbf{k})$.
Since the spin dependence has been considered in details in Sec.~\ref{sec:NNinter}, here we focus on the spin independent part of the tight biding Hamiltonian. 
As we will see, the generic Hamiltonian describes hopping processes into nearest and next to nearest neighboring sites.
The hopping amplitudes for these processes are defined similarly to Eq.~\eqref{eq:T_BA}, 
\begin{align}\label{eq:T_A(B)}
    T_{A}(\mathbf{R}^{n(nn)}) = \frac{1}{2} \sum_s \langle \phi_{A}( \mathbf{r} - \mathbf{R}^{n(nn)}) \chi_{s} | H |\phi_{A}(\mathbf{r}) \chi_{s}  \rangle\, ,
\end{align}
where the $\mathbf{R}^{n}$ and $\mathbf{R}^{nn}$ denote the vectors connecting nearest and next to nearest neighbours on the $A$ sublattice.
The same definition is adopted for  $T_{B}(\mathbf{R}^{n(nn)})$ for the hopping amplitude on the $B$ sublattice.
The vector $\mathbf{R}^{n}$ can take six values, $\pm \hat{x}$, $\pm \hat{y}$ and $\pm \hat{z}$.
The vector $\mathbf{R}^{nn}$ can take twelve values, $\pm \hat{x}$, $\pm \hat{y}$ and $\pm \hat{z}$, $m_1 \hat{x} + m_2 \hat{y}$, $m_1 \hat{y} + m_2 \hat{z}$ and $m_1 \hat{x} + m_2 \hat{z}$ for $m_{1,2} = \pm 1$.

Consider first the nearest neighbor hopping amplitudes, $\hat{H}^a_n(\mathbf{k})$.
Both at finite and zero SOI, the symmetry constrains it to be the same for the two sublattices, 
\begin{align}\label{eq:intra_H}
    \hat{H}^a_n(\mathbf{k}) &=  \varkappa_0 \sigma_0\left[A_0 + f_{1}(\mathbf{k})\right],
    \notag \\
    f_{1}(\mathbf{k}) &= A_x \cos k_x + A_y \cos k_y + A_z \cos k_z,
\end{align}
where $A_0$, $A_x$, $A_y$ and $A_z$ are three real parameters.

Qualitatively, we expect a sublattice dependence at the level of the next to nearest neighbors, see Fig.~\ref{fig:orbitals}b. 
In this case the symmetry reduces the twenty four real parameters down to four, see Tab.~\ref{tab:tnnn}.
Notably, $T_A(\hat{x}\pm\hat{y}) = A_{\mp}$, $T_B(\hat{x}\pm\hat{y}) = A_{\pm}$, with all the rest of the amplitudes being equal on the two sublattices.
The feature essential for the altermagnetism is $A_+ \neq A_-$.
Since $\hat{H}^a_{nn}(\mathbf{k})$ is generally weaker than $\hat{H}^a_{n}(\mathbf{k})$ it is permissible 
to retain only the essential part of $\hat{H}^a_{nn}(\mathbf{k})$ by setting, $A_+ = - A_- = A'$, and the rest of the amplitudes to zero,
\begin{align}\label{eq:Ha_nn}
    \hat{H}^a_{nn}(\mathbf{k}) = 2 A' \varkappa_z \sigma_0 f_2 (\mathbf{k})\, , \, f_2 (\mathbf{k}) = 2 \sin k_x \sin k_y\, .
\end{align}

Qualitatively, the emergence of a finite spin splitting at zero SOI can be understood as originating from the difference in the orbital wave function for the two sublattices as shown in Fig.~\ref{fig:orbitals}b.
We note, however, that even if such difference is not included, the finite $A'$ in Eq.~\eqref{eq:Ha_nn} results from the sublattice dependent potential experienced by an electron hopping to the next to nearest neighbours in the $xy$-plane.
Indeed, this has been found in Ref.~\cite{Roig2024} for a model with identical $d_{x^2 - y^2}$ orbital wave function on the two sublattices.
In fact, these are the two manifestations of the same sublattice asymmetry. 

\subsection{Band degeneracies of the model Hamiltonian}
Here we discuss the band degeneracies of the model Hamiltonian, Eq.~\eqref{eq:Htotal}, and compare them with the general results of Sec.~\ref{sec:Symmetries}.
\subsubsection{Non-zero SOI}
In this case the Hamiltonian, Eq.~\eqref{eq:Htotal}, is given by Eqs.~\eqref{eq:HeT}, \eqref{eq:HeB}, \eqref{eq:intra_H} and \eqref{eq:Ha_nn}.
We have confirmed the double degeneracy at $\mathbf{k} \in \mathcal{K}_{\mathrm{SO}}$.

\subsubsection{Zero SOI}
To study the spectrum degeneracies it is enough to consider the difference, 
$\Delta \hat{H}(\mathbf{k}) =\hat{H}(\mathbf{k}) -\hat{H}^a_{n}(\mathbf{k})$, as the term $\hat{H}_n^a(\mathbf{k}) \propto \varkappa_0 \sigma_0$, Eq.~\eqref{eq:intra_H}.
According to Eq.~\eqref{eq:He_noSO}, the spin along the magnetization is a good quantum number, and the spectrum splits into two bands $ \Delta \hat{H}_\pm$ for the two spin wave functions, 
$\bar{\chi}_\pm =  ( \chi_{\uparrow} \pm i \chi_{\downarrow})/\sqrt{2}$ satisfying 
$\sigma_y \bar{\chi}_{\pm} = \pm \bar{\chi}_\pm$, 
\begin{align}\label{eq:Htotal_noSO1}
  \Delta \hat{H}_\pm (\mathbf{k})& = \pm B_{ex} \varkappa_z +  2 A' \varkappa_z  f_2 (\mathbf{k})  
  \notag \\
& +  8 t_0^R \varkappa_x  \cos\frac{k_x}{2} \cos\frac{k_y}{2} \cos\frac{k_z}{2}  
    \notag \\
 & \mp    8 t_y^R \varkappa_x  \sin\frac{k_x}{2} \sin\frac{k_y}{2} \cos\frac{k_z}{2} \, .
\end{align}
The eigenvalues $E_\pm$ of $\Delta \hat{H}_\pm$ satisfy,
\begin{align}\label{eq:Epm}
    E_\pm^2 &  =  [B_{ex} \pm 2 A'  f_2 (\mathbf{k})]^2 
    +
    64 \cos^2\frac{k_z}{2}
    \notag \\
    & \times 
    \left(\! t_0^R \cos\frac{k_x}{2} \cos\frac{k_y}{2} \pm   t_y^R   \sin\frac{k_x}{2} \sin\frac{k_y}{2}\! \right)^2.
\end{align}
The spin degeneracy amounts to the condition, $E_+^2 = E_-^2$.
This implies with necessity, $f_2 (\mathbf{k})=0$, namely $\mathbf{k} \in \mathcal{K}_{\mathrm{alt}}$.
If $k_x =0$ or $k_y=0$, the last term in the second line of Eq.~\eqref{eq:Epm} vanishes.
If $k_x =\pi$ or $k_y=\pi$ the first term does.
In both cases we indeed have a degeneracy only for $\mathbf{k} \in \mathcal{K}_{\mathrm{alt}}$ as expected.

\begin{figure*}[t!]
\begin{center}
\centering
\includegraphics[width=1.0\textwidth]{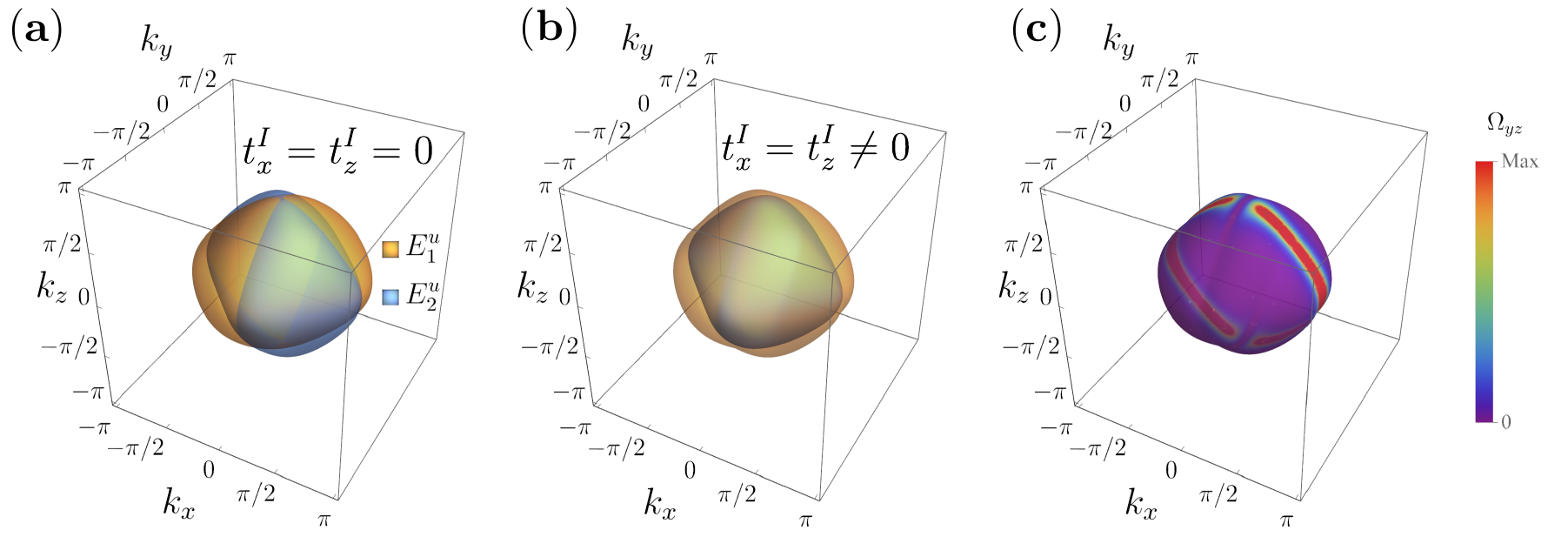}
\caption{
Fermi surface for the effective Hamiltonian, \( \hat{H}_{u} (\mathbf{k}) \), Eq.~\eqref{eq:Heffective}.
(a) At zero SOI, \(t_x^I = t_z^I\), Eq.~\eqref{eq:h_effective} the two bands are degenerate 
at $k_x=0$, $k_y=0$ planes. 
(b) SOI splits the bands.
(c) Berry curvature \( \Omega^u_{yz} \), Eq.\eqref{eq:Omega_yz} computed on the external Fermi surface, peaks at the \(  k_y = 0 \) meridian. 
} \label{fig:Fermi_wide}
\end{center}
\end{figure*}


\section{Anomalous Hall effect}
\label{sec:AHE}

The AHE can be understood based on the symmetry considerations.
The AHE is finite as $\mathcal{T}$ symmetry is replaced by the combined $\mathcal{T}C_{2z}$ operation.
Due to the Onsager relation, $\mathcal{T}C_{2z}$ imposes the restrictions, 
$\hat{\sigma}_{xy} = \hat{\sigma}_{yx}$, $\hat{\sigma}_{xz} = - \hat{\sigma}_{zx}$ and $\hat{\sigma}_{yz} = - \hat{\sigma}_{zy}$ on the conductivity tensor, $\hat{\sigma}$.
This implies at most the transport anisotropy in the $xy$-plane in the form of the planar Hall effect, yet
no AHE as \(\sigma^H_{xy} = (\sigma_{xy} - \sigma_{yx})/2 =0\).
In contrast to $\mathcal{T}$, the combined $\mathcal{T}C_{2z}$  symmetry is consistent with $\sigma^H_{xz} \neq 0$ and $\sigma^H_{yz} \neq 0$.
The unitary $\boldsymbol{\tau}C_{2x}$ symmetry imposes $\sigma^H_{xz} = 0$ and still allows a finite $\sigma^H_{yz}$.

An alternative way to see this is to notice that the ferromagnetic exchange field $\mathbf{B}_F = B_F \hat{x}$ is consistent with the magnetic point group.
And the symmetry that allows for a finite $\mathbf{B}_F$ also allows for a finite $\sigma^H_{yz}$.
It turns out that the weak ferromagnetic component causes a slight enhancement of the AHE, (see App.~\ref{SI:canting} for details). 

Physically the finite ferromagnetic exchange field can be thought of as resulting from the canting of the antiferromagnetic magnetization. 
Such canting must originate from the SOI.
Indeed, in this case the spin-only group is inconsistent with a finite $\mathbf{B}_F$.
We expect, therefore that $\hat{\sigma}_{yz}=0$ vanishes at zero SOI. 
Indeed, in this case the additional $C_{2z}$ rotation symmetry applied solely to the orbital motion with spins left untouched enforces $\hat{\sigma}_{yz}=0$ as the current operator is spin independent.
Therefore, a finite SOI is required for AHE.
This is unlike the $\pi$-transition in altermagnet Josephson junctions \cite{Ouassou2023}.

The intrinsic contribution to AHE \cite{Sinitsyn2007,Xiao2010},
\begin{align}\label{eq:sigma_H}
    \sigma^H_{ij} = e^2 \sum'_{b \mathbf{k} } \Omega^b_{ij}(\mathbf{k})
\end{align}
relates it to the Berry curvature antisymmetric tensor \( \Omega^b_{ij}(\mathbf{k}) = - 2 \mathrm{Im}\langle \partial_{k_i} u_b(\mathbf{k}) | \partial_{k_j} u_b(\mathbf{k} ) \rangle \), where 
$u_b(\mathbf{k})$ are periodic parts of the Bloch functions, at the band $b$ and momentum $\mathbf{k}$.
At zero temperature the primed summation in Eq.~\eqref{eq:sigma_H} runs over all occupied Bloch states.

Here we limit the consideration to the large exchange field.
In this case the upper ($u$) and lower ($l$) band doublets are clustered around $+B_{ex}$ and $-B_{ex}$, respectively.
The two band doublets belong to the upper ($V_{u}$)  and lower ($V_l$)  subspaces,  
$V_{u(l)} = \{\phi_{A}(\mathbf{r}) \bar{\chi}_{+(-)}, \phi_{B}(\mathbf{r}) \bar{\chi}_{-(+)} \}$ that 
approximately decouple.

Projecting the total Hamiltonian, Eq.~\eqref{eq:Htotal} onto $V_{u(l)}$ spaces we obtain two simpler two by two effective Hamiltonians,
\begin{align}\label{eq:Heffective}
    \hat{H}^{u(l)} (\mathbf{k}) = \pm B_{ex}\rho_0  + f_1(\mathbf{k})\rho_0 - \mathbf{h}^{u(l)}(\mathbf{k}) \cdot \boldsymbol{\rho} \, ,
\end{align}
where $\rho_0$ and $\boldsymbol{\rho}$ are the unit and Pauli pseudo-spin matrices operating in $V_\pm$, and
\begin{align}\label{eq:h_effective}
    h^{u(l)}_x & =  \mp 8 t_x^I \sin\frac{k_x}{2} \cos\frac{k_y}{2} \sin\frac{k_z}{2},  
    \notag \\
    h^{u(l)}_y & =  - 8 t_z^I \cos\frac{k_x}{2} \cos\frac{k_y}{2} \cos\frac{k_z}{2},
    \notag \\
    h^{u(l)}_z & = -2 A' f_2(\mathbf{k})\, ,
\end{align}
where we ignore the contribution of the $\hat{H}_{\mathrm{B}}^{e}(\mathbf{k})$, Eq.~\eqref{eq:HeB} for brevity.
The $x$-component of $\mathbf{h}$ is opposite at the upper and lower subbands.
This ensures  that AHE vanishes when all the bands are occupied, and the system is a topologically trivial insulator.
Here we consider the metallic regime with the upper (lower) band doublets partially (fully) occupied, respectively. 
This situation is described by $\hat{H}_{u} (\mathbf{k})$.
It gives rise to the two Fermi surfaces split by the spin dependent terms of Eq.~\eqref{eq:Heffective}, see Fig.~\ref{fig:Fermi_wide}.
Even though the spectrum appears as symmetric under the $C_{2z}$ operation, it is not a symmetry at finite SOI.
Indeed, Eq.~\eqref{eq:h_effective} shows that $h^{u(l)}_x$ flips under this operation.
Therefore Bloch states have a different spin dependence at momenta related by $C_{2z}$.

The integration region in Eq.~\eqref{eq:sigma_H} is contained in between the two split Fermi surfaces arising from $\hat{H}_{u} (\mathbf{k})$.
The straightforward calculation based on Eqs.~\eqref{eq:Heffective} and  \eqref{eq:h_effective} yields $\sigma^H_{xy} =0$, since $\Omega^u_{xy}(-k_x,k_y,k_z) = -\Omega^u_{xy}(k_x,k_y,k_z)$. 
Similarly, $\sigma^H_{xz}=0$.
The nonzero $\sigma^H_{yz}$ arises from
\begin{align}\label{eq:Omega_yz}
\Omega^u_{yz}(\mathbf{k}) = 
\frac{A' t_x^I t_z^I [2 \cos(k_y/2)]^4 (\sin k_x)^2}{|h^u(\mathbf{k})|^{3}}\, .
\end{align}

We can see that in the limit of the weak SOI, the $k_y=0$ plane hosts the nodal line where $h^u(\mathbf{k})=0$, and Eq.~\eqref{eq:Omega_yz} becomes non-analytic.
It is natural to expect this degeneracy to produce the non-analytic dependence of the AHE on SOI. 
To clarify this point we consider the set of parameters resulting in a small Fermi surface centered at the $\Gamma$ point, see Fig.~\ref{fig:Fermi_wide}a,b.
We take for the diagonal part of the effective Hamiltonian \eqref{eq:Heffective}, 
$f_1(\mathbf{k}) = E_0 \mathbf{k}^2$, which obtains by setting $A_0 = 6 E_0$, and $A_x = A_y = A_z = -2 E_0$ in Eq.~\eqref{eq:intra_H}.
We next set the chemical potential counted relative to the bottom of the band at $B_{ex}$, $\mu \ll E_0$. 
The Fermi momentum becomes $k_F = \sqrt{\mu/E_0} \ll 1$.
In the same limit of a small and nearly spherical Fermi surface we approximate
\begin{align}\label{eq:h_xyz}
    h^u_x  \approx   - 2 t_x^I k_x k_z  , \,  \,\,  \,  \,\,
    h^u_y  \approx  -8  t_z^I,\,\,\,  \, \,\,
    h^u_z  \approx - 2 A' k_x k_y.
\end{align}
With these approximation, taking into account the smallness of the Fermi energy, and setting $t_x^I = t_z^I = t$ for the SOI, the Berry curvature simplifies to 
\begin{align}\label{eq:Omega_yz1}
\Omega^u_{yz}(\mathbf{k}) \approx 
\frac{2 A' k_x^2 t^2}{\left(A'^2 k_x^2 k_y^2+(4t)^2\right)^{3/2}}\, .
\end{align}

To find the asymptotic behavior of $\sigma^H_{yz}$ in the limit $t \rightarrow 0$ note that the Berry curvature is strongly localized in the $k_y=0$ plane, see Fig.~\ref{fig:Fermi_wide}c.
For the fixed $k_x$ and $k_z = \sqrt{k_F^2 - k_x^2}$ on a $t=0$ Fermi surface the $k_y$ integration range, $|k_y|< \bar{k}_y$ is fixed by the condition, $E_0 \bar{k}_y^2 - \left(A'^2  k_x^2 \bar{k}_y^2 + (4t)^2\right)^{1/2} <0$.
This gives $\bar{k}_y = A' |k_x|/E_0$ not too close to the north pole of the Fermi sphere,
$\bar{k}_x < |k_x|<k_F $, with $\bar{k}_x = 4\sqrt{|t| E_0}/A'$.
Except for a tiny interval of $|k_x|< 2 \bar{k}_x$ the $k_y$ integration of $\Omega^u_{yz}(\mathbf{k})$ converges fast enough to approximate 
\begin{align}\label{eq:sigma_yz}
\sigma^H_{yz} \approx 
4 \int_{-k_F}^{-\bar{k}_x}  \frac{ d k_x}{2 \pi} \int_{ \bar{k}^{(1)}_z}^{\bar{k}^{(2)}_z} \frac{d k_z}{2 \pi} 
\int_{-\infty}^{\infty}  \frac{ d k_y}{2 \pi}
\Omega^u_{yz}(\mathbf{k}) \, ,
\end{align}
where the factor of four accounts for the contributions of the four sectors in the $k_y=0$ meridian of the Fermi sphere.
The straightforward integration over $k_y$ results in 
\begin{align}\label{eq:sigma_yz1}
\sigma^H_{yz} \approx \frac{1}{(2 \pi)^3} 
\int_{-k_F}^{-\bar{k}_x}  d k_x |k_x| \int_{ \bar{k}^{(1)}_z}^{\bar{k}^{(2)}_z} d k_z\, .
\end{align}
In the small $t$ regime, the limits of the $k_z$ integration are set by the energy split $2 |\mathbf{h}(\mathbf{k})| \approx  16 |t|$ in the $k_y=0$ plane, see Eq.~\eqref{eq:h_xyz}. 
This allows us to perform the $k_z$ integration in Eq.~\eqref{eq:sigma_yz1}, and write 
\begin{align}\label{eq:sigma_yz2}
\sigma^H_{yz} \approx \frac{2|t|}{v_F \pi^3} 
\int_{-k_F}^{0}  d k_x \frac{|k_x|}{k_F}\sqrt{k_F^2 - k_x^2} = 
 \frac{2 |t| k_F^2}{3 v_F \pi^3} 
,
\end{align}
where the Fermi velocity, $v_F = 2 E_0 k_F$, and we have set the upper limit of the $k_x$ integration, $- \bar{k}_x$ to zero, as the contribution of the region close to the north pole is negligible in this limit.

\begin{figure}
\begin{center}
\centering
\includegraphics[width=0.47\textwidth]{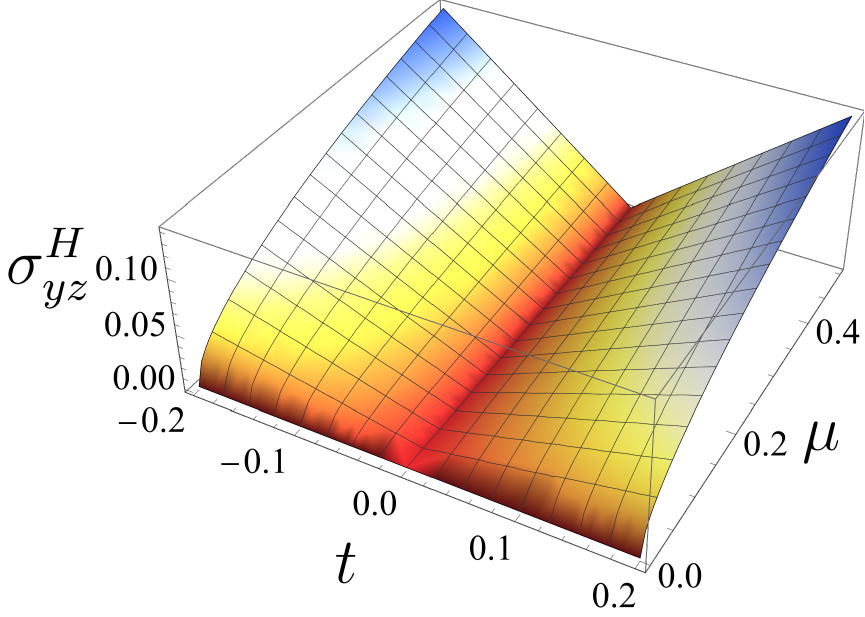}
\caption{
Hall conductivity, \(\sigma^H_{yz}\) as given by Eq.~\eqref{eq:sigma_yz2} as a function of spin-orbit interaction strength, \(t\), and Fermi Energy, \(\mu\), valid in the regime where the altermagnetic splitting exceeds the SOI spin splitting (arbitrary units).
} \label{fig:Hall}
\end{center}
\end{figure}

The numerical coefficient in Eq.~\eqref{eq:sigma_yz2}  depends on the details of the model.
The generic feature of Eq.~\eqref{eq:sigma_yz2} is the non-analytic linear dependence of the Hall conductivity on the SOI.
We have traced it's origin to the crossing lines of the Fermi surface(s) with the plane $k_y =0$.
This non-analyticity is present whenever such crossing occurs, and in this sense is universal. 
The concentration of the Berry curvature at these crossing(s) expressed via Eq.~\eqref{eq:Omega_yz1} has been reported for the specific model of SOI in Ref.~\cite{Smejkal2020}.
The linear in SOI Hall conductivity agrees with the more recent numerical calculations \cite{Roig2024}.
At the same time the result, Eq.~\eqref{eq:sigma_yz2} vanishes in the insulating phase as the Fermi surface shrinks.
Both trends are illustrated in the Fig.~\ref{fig:Hall}.

At weak SOI the Hall conductivity, Eq.~\eqref{eq:sigma_yz2} is independent of the exchange splitting $A'$.
It holds for $t < \mu (A'/E_0)^2$ which may be not too restrictive given that the band width and the Fermi energy, $\mu$ are of the same order and $A'$ is few tens of meV based on the DFT of Ref.~\cite{Smejkal2023} performed on RuO$_2$.


\section{Summary and Outlook}
\label{sec:Summary}

In this work, we have computed the intrinsic anomalous Hall effect of an altermagnet based on the minimal symmetry-constrained tight-binding model.
We have studied the collinear altermagnet with and without SOI separately. 
At finite SOI we have employed the analysis based on the magnetic group symmetry of FeSb$_2$.
We have found the terms that are either even or odd under the $\mathcal{T}$ symmetry. 
The terms belonging to the first category agree with the terms obtained in the approaches based on the non-magnetic space groups \cite{Roig2024,Antonenko2024}.
The terms odd in $\mathcal{T}$ describe a combined action of the altermagnetic order parameter and SOI.
These require the magnetic group approach.

We next analysed the zero SOI in the framework of the spin symmetries.
In this case we have identified and presented explicitly the symmetry elements including the possibility of acting differently on spin and orbital degrees of freedom.
The tight binding model in this case is considerably more restricted compared to the case of a finite SOI.
We note that both the elements of spin-only as well as the non-trivial elements of spin group are instrumental in restricting the zero SOI model.

Comparison of the models bases on magnetic and spin groups respectively allows one to single out the effect of the SOI on the dispersion relation.
This would be hard to achieve without having done both types of analysis.
We have worked out the band degeneracies based on the Wigner criterion. 
Usually it is done for the magnetic or non-magnetic space groups.
We have extended this treatment to the spin group in question.
The summation over the spin-only elements in this case has to be understood as integration over the continues spin rotation angle.
The tight binding models in both the finite and zero SOI comply with the general symmetry requirements.
And we have used the properly generalized Wigner criterion to benchmark our results.

Here we have focused on FeSb$_2$ material, yet our conclusions are qualitatively similar for other altermagnetic materials such as RuO$_2$.
According to Refs.~\cite{Roig2024,Antonenko2024} the SOI has a similar form in two materials.
Since the exchange induced splitting has in fact the same form er expect the same results apply qualitatively to both types of systems.

This work addresses the intrinsic contribution to AHE. 
Our model is general, and yet simple enough to serve as a starting point to explore the extrinsic contributions to AHE.
One can expect the extrinsic contribution to dominate in the metallic regime where the Fermi energy exceeds the gap due to the SOI as well as the gap induced by the exchange interaction.
This task is relegated for future studies.


\section*{Acknowledgements}

We would like to thank I. Mazin for invaluable discussions that have been a source of inspiration for the current work.
We are grateful to D. Agterberg and B. Andersen for sharing with us their recent results and pointing our attention to an unusual linear dependence of the Hall conductivity in the SOI.
We are grateful to D. Antonenko for providing us with the explanation of their recent work.
We thank R. Lifshitz for the discussion of magnetic space groups.
We thank I. Gruzberg for the discussion of the time reversal symmetry as acting on degrees of freedom spin and orbital motion included.
We thank M. Smolkin for pointing our attention to universality of the time reversal symmetry, independent on the Lorentz boost magnitude. 
L. A. and M. K. acknowledge the financial support from the Israel Science Foundation, Grant No.~2665/20. A. L. acknowledges the
financial support by the National Science Foundation Grant No. DMR-2203411 and H. I. Romnes Faculty Fellowship provided by the University of Wisconsin-Madison Office of the Vice Chancellor for Research and Graduate Education with funding from the Wisconsin Alumni Research Foundation. 


\appendix
\section{Anti-unitary spin symmetries}
\label{SI:A}

Here we show that the anti-unitary spin symmetries cause no additional degeneracy.
The elements of the spin group are listed in Eq.~\eqref{SpG_60}. 
In Sec.~\ref{sec:Zero_SOI} the unitary half of the spin point group, Eq.~\eqref{eq:SG_13} has been shown to 
produce the double degeneracy at momenta, $\mathbf{k} \in \mathcal{K}_{alt}$, where $\mathcal{K}_{alt}$ contains four planes, $k_x=0$, $k_y=0$, $k_x=\pi$, $k_y=\pi$.

At a given $\mathbf{k}$, the degeneracy doubles due to the anti-unitary operations if at least one of the anti-unitary operators in Eq.~\eqref{SpG_60} with $\mathcal{T}$ removed flips $\mathbf{k}$.
In this case, the extra degeneracy appears in cases (b) and (c), 
\begin{align}\label{eq:W_test_SG}
    \sum'_{g \in \mathcal{G}^u_S}\chi_{\mathbf{k}}\left\{ [\boldsymbol{\tau}(E||C_{2x}) g]^2 \right\} = \begin{cases}
    + [\mathbf{k}]\mathcal{T}^2  & \mathrm{case \,(a)} \\
    -[\mathbf{k}]\mathcal{T}^2  & \mathrm{case \,(b)} \\
    0 & \mathrm{case \,(c)}
    \end{cases}\, .
\end{align}

The symmorphic unitary operators include the rotations around $\hat{y}$, $C_{y}(\varphi)$ by an arbitrary angle, $\varphi$.
For $\mathbf{k} \notin \mathcal{K}_{alt}$, $\chi_{\mathbf{k}}[C_{y}^2(\varphi)] = \exp( \mp i \varphi)$, and for $\mathbf{k} \in \mathcal{K}_{alt}$, $\chi_{\mathbf{k}}[C_{y}^2(\varphi)] = 2 \cos \varphi$.
In both cases $\int_0^{2 \pi} d \varphi/(2 \pi)\chi_{\mathbf{k}}[C_{y}^2(\varphi)] = 0$ and therefore the symmorphic operations do not contribute to the sum.

In contrast the nonsymmorphic unitary operations do contribute, since
$[C_{2x}C_{y}(\varphi)]^2 = \bar{E}$, and therefore 
$\chi_{\mathbf{k}}[C_{2x}C_{y}(\varphi)]^2 = -1 (-2)$ for $\mathbf{k} \notin \mathcal{K}_{\mathrm{alt}}$,  ($\mathbf{k} \in \mathcal{K}_{\mathrm{alt}}$).
The action of symmorphic operations on the momentum, $\mathbf{k}$ is the same as that of the four operations, $\left\{ E  , \mathcal{P} , m_z , C_{2z}  \right\} $ given in the Tab.~\ref{tab:SG2}.

\begin{table}[h]
\caption{\label{tab:SG2} 
The action of the orbital part of the operations from the list $\left\{ E  , \mathcal{P} , m_z , C_{2z}  \right\} $ on the momentum $\mathbf{k}=(k_x,k_y,k_z)$.}
\begin{tabular}{|c|c|c|c|}
\hline
$E$ & $\mathcal{P}$ & $m_z$ & $C_{2z}$ 
 \\
 \hline
 $(k_x, k_y, -k_z)$ & $(-k_x, -k_y, -k_z)$   & $(k_x, k_y, -k_z)$   & $(-k_x, -k_y, k_z)$  
\\
\hline
\end{tabular}
\end{table}

For a generic $\mathbf{k}$ only the inversion $\mathcal{P}$ contributes to the sum, Eq.~\eqref{eq:W_test_SG}, and we have case (a).
At the $\Gamma$-point, $\mathbf{k}=0$ all eight unitary operation reverse $\mathbf{k}$ such that $[\mathbf{k}]=8$.
And the sum is the product of the point group character $-2$ and the number of elements contributing $+4$, and we have case (a) again.
Similarly one can show that the case (a) holds throughout the Brillouin Zone, and indeed there is no degeneracy caused by the anti-unitary operations. 
So we get no degeneracy doubling due to the anti-unitary operations in the case of a given spin symmetry group.

\section{Symmetry constrained hopping amplitudes}
\label{SI:amplitudes}
\begin{figure}
\begin{center}
\centering
\includegraphics[width=0.47\textwidth]{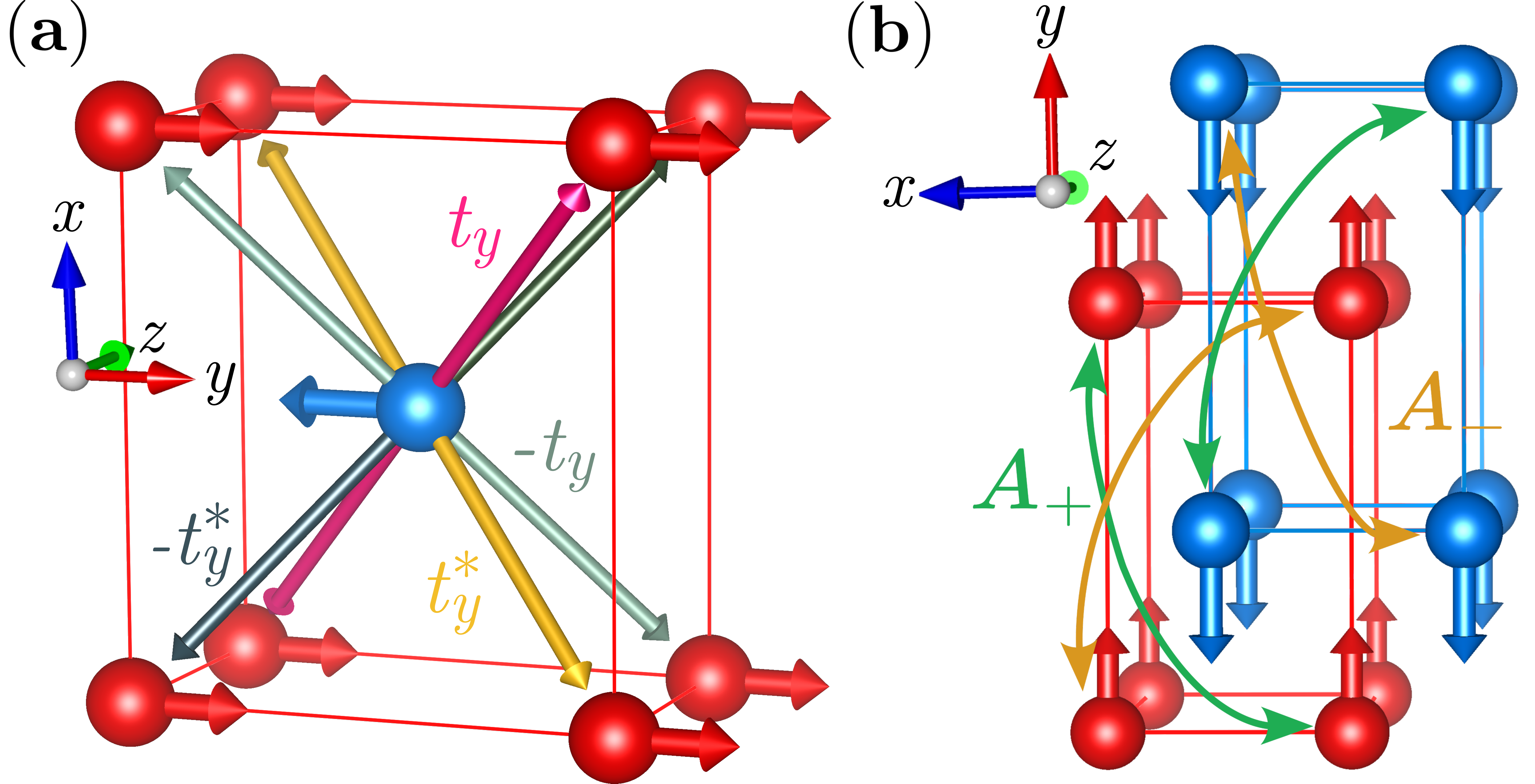}
\caption{
Amplitudes of hopping between the Fe sites.
(a) Inter-sublattice nearest neighbor hopping amplitudes as summarized in Tab.~\ref{tab:t}. 
Only the terms of Eq.~\eqref{eq:T_BA_1} \( \propto \sigma_y \) are displayed.
(b) Inra-sublattice next-nearest neighbor hopping amplitudes.
Only the processes that are distinct for the two sublattices are shown, with the full list of amplitudes given by Tab.~\ref{tab:tnnn}.
} \label{fig:hopping_amp}
\end{center}
\end{figure}

Here we illustrate the constrains imposed on hopping amplitudes by the magnetic group symmetry, Fig.~\ref{fig:hopping_amp}.
The most general form of the inter-sublattice nearest neighbor hopping amplitudes is given in Tab.~\ref{tab:t}. 
Here we illustrate the constrains contained in this table in Fig.~\ref{fig:hopping_amp}a.
For clarity, we do it just for one out of four types of hopping amplitudes presented in Eq.~\eqref{eq:T_BA_1}.
Namely we consider the $\mu =y$ term of Eq.~\eqref{eq:T_BA_1} and show the hopping amplitudes in Fig.~\ref{fig:hopping_amp}a based on Tab.~\ref{tab:t}. 
The remaining three amplitudes can be illustrated in a very similar way based on the same table.

The full information on the intra-sublattice hopping amplitudes, $T_{A(B)}(\Delta \mathbf{R})$ over the distance $\mathbf{R}$ is contained in Tab.~\ref{tab:tnnn}.
Here we show the intra-sublattice hopping amplitudes important for an altermagnetism in Fig.~\ref{fig:hopping_amp}b.

\section{The effect of magnetization canting on AHE}
\label{SI:canting}
Here we study the effect of  a weak ferromagnetic component of magnetization on AHE.
To clarify this we consider the limit where the exchange field $\mathbf{B}_x = B_x \hat{x}$ is added to the collinear staggered magnetization, such that $B_x \ll B_{ex}$.
We do not intend to cover all possible cases, and instead consider the limit of the $B_{x}$ exceeding the spin splitting at $B_{x}=0$.

To find the small correction to AHE in this case we employ the method of effective Hamiltonian \cite{Cohen1992}.
It is a generalization of the standard perturbation theory to the case of quasi-degenerate bands.
In our problem we have two quasi-degenerate spaces, 
$V_{u(l)} = \{\phi_{A}(\mathbf{r}) \bar{\chi}_{+(-)}, \phi_{B}(\mathbf{r}) \bar{\chi}_{-(+)} \}$ separated by the large exchange energy splitting.
In the four dimensional space spanned by the four states $\{\phi_{A}(\mathbf{r}) \bar{\chi}_{+}, \phi_{B}(\mathbf{r}) \bar{\chi}_{-},\phi_{A}(\mathbf{r}) \bar{\chi}_{-}, \phi_{B}(\mathbf{r}) \bar{\chi}_{+} \}$ the Hamiltonian takes the form,
\begin{align}\label{eq:Bx_13}
    H = \begin{pmatrix}
        \hat{H}^u & \hat{V}_{u l } \\
        \hat{V}_{ l u }  & \hat{H}^l
    \end{pmatrix}\, ,
\end{align}
where the block diagonal part is specified by Eqs.~\eqref{eq:Heffective} and \eqref{eq:h_effective}. 
For the exchange field, $\mathbf{B}_x = B_x \hat{x}$ we have 
\(\hat{V}_{u l } = - i \rho_z \), and \(\hat{V}_{l u } = \hat{V}_{u l }^\dagger \).

The coupling $\hat{V}_{u l }$ affects the energy levels as well as wave functions of the problem defined for the two-dimensional upper subspace $V_{u}$.
To the second order in $B_x/B_{ex}$, the energy levels are fixed by the effective Hamiltonian, 
$\hat{H}^{eff} = \hat{H}^u + \Delta \hat{H}^{eff}$, with 
\begin{align}\label{eq:Bx_15}
   \Delta \hat{H}^{eff}_{ss'} = &  \frac{1}{2}\sum_{k} {}^u \langle s | \hat{V}_{u l }| k \rangle^l 
    \notag \\
    \times & \left( \frac{1}{E^u_s - E^l_k} + \frac{1}{E^u_{s'} - E^l_k} \right){}^l\langle k | \hat{V}_{ l u } | s'  \rangle^u \, ,
\end{align}
where the states $\hat{H}^{u(l)} | s  \rangle^{u(l)} = E^{u(l)}_s | s  \rangle^{u(d)}$, $s=1,2$ diagonalize the two decoupled Hamiltonians.
Based on Eq.~\eqref{eq:Heffective} the energies of the decoupled Hamiltonians read
\begin{align}\label{eq:Bx_17}
    E^{u(l)}_{s=1}(\mathbf{k})  & = \pm B_{ex} + f_1(\mathbf{k}) + |\mathbf{h}^{u(l)}(\mathbf{k}) |
    \notag \\
    E^{u(l)}_{s=2}(\mathbf{k})  & = \pm B_{ex} + f_1(\mathbf{k}) - |\mathbf{h}^{u(l)}(\mathbf{k}) |\, .
\end{align}
We compute the correction, $\Delta H^{eff}$ to the effective Hamiltonian given by Eq.~ \eqref{eq:Bx_15} to the first order in $h/B_{ex} \ll 1$.
The first contribution, $\Delta H^{eff}_{(a)}$ comes from the expansion of $E_{k}^l$, and setting $E_s^u$ to $B_{ex}$. 
This contribution takes the form in the original basis $V_u$,
\begin{align}\label{eq:Bx_19}
    \Delta \hat{H}^{eff}_{(a)} = -\frac{B^2_x }{4 B^2_{ex}} \rho_z \left[\mathbf{h}^l(\mathbf{k})\cdot \boldsymbol{\rho} \right]\rho_z\, .
\end{align}
Additional contribution $\Delta H^{eff}_{(b)}$ originates from the expansion of $E_{s,s'}^{u(l)}$ in $h^u/B_{ex}$ and setting $E_{k}^{l}$ to $-B_{ex}$ in Eq.~\eqref{eq:Bx_15},
\begin{align}\label{eq:Bx_21}
    \Delta \hat{H}^{eff}_{(b)} = - \frac{B^2_x }{4 B^2_{ex}}  \left[\mathbf{h}^u(\mathbf{k})\cdot \boldsymbol{\rho} \right]\, .
\end{align}
Combining Eqs.~\eqref{eq:Bx_19} and \eqref{eq:Bx_21} we obtain for the total correction, 
\begin{align}\label{eq:Bx_23}
    \Delta \hat{H}^{eff} =  -\frac{B^2_x }{2 B^2_{ex}} h^u_y(\mathbf{k})  \rho_y \, .
\end{align}
The result \eqref{eq:Bx_23} indicates that the effect of the ferromagnetic magnetization component in the considered limit is the renormalization of the $t_z^I$ interaction amplitude as defined in Eq.~\eqref{eq:h_effective} to the effective one, 
\begin{align}\label{eq:Bx_25}
    t_z^{I,eff} = t_z^I\left(1 + \frac{B_x^2}{2 B_{ex}^2} \right) \, .
\end{align}
Here we focused on the spectrum renormalization. 
In fact the wave functions also change as a result of the perturbation.
This effect can be shown to be negligible in the considered range of parameters.

The Eq.~\eqref{eq:Bx_25} indicates that the weak ferromagnetism causes a a slight enhancement of the AHE.


%

\end{document}